# The Rotational Light Curve of (79360) Sila-Nunam, an Eclipsing Binary in the Kuiper Belt

David L. Rabinowitz[1], Susan D. Benecchi[2], William M. Grundy[3], Anne J. Verbiscer[4]

[1]Yale University, Center for Astronomy and Astrophysics, P.O. Box 208120, New Haven, CT 06520-8120, United States, david.rabinowitz@yale.edu 203 432 3391

[2]Planetary Science Institute, 1700 E. Fort Lowell, Suite #106, Tucson, AZ 85719, United States

[3]Lowell Observatory, 1400 W. Mars Hill Road, Flagstaff, AZ 86001, United States

[4]University of Virginia, Department of Astronomy, PO Box 400325, Charlottesville, VA 22904, United States

## Abstract

We combine long-term photometric observations in multiple band passes to determine the rotational light curve for the binary Kuiper-Belt object (79360) Sila-Nunam. We measure an unambiguous fundamental period of 6.2562 ± 0.002 d, within 0.02% of half the orbital period ($P_{orb}$ = 12.50995 ± 0.00036 d) determined earlier from HST observations resolving the binary. The light curve is double-peaked, and well fit by the sum of two sinusoids: a primary with period $P_{orb}$/2 and peak-to-peak amplitude 0.120 ± 0.012 mag and a secondary with period $P_{orb}$ and peak-to-peak amplitude 0.044 ± 0.010 mag. Excluding observations within ~0.1 deg of opposition, we measure a linear solar phase dependence with slope 0.147 ± 0.018 mag deg$^{-1}$ and a mean absolute magnitude in the Gunn g band of 6.100 ± 0.006 mag. There is no rotational color variation exceeding 4%. We also observe that eclipses occur centered on light curve minima to within 0.3%, requiring the long axis of at least one of the two bodies to point precisely toward the other. Assuming the binary is doubly synchronous and both rotation axes are aligned with the orbital angular momentum vector, our observations jointly constrain triaxial shape models for Sila and Nunam such that the product of their long-to-intermediate axes ratios is 1.120 ± 0.01. Hence both bodies are elongated by 6%, or else one is elongated by 6% to 12%, and the other by less than 6%.

## 1. Introduction

The binary object (79360) Sila-Nunam is one of the most well studied bodies beyond the orbit of Neptune. Since its discovery by Luu et al (1997), this Kuiper-Belt object (KBO) has been the frequent target of photometric observations (Romanishin & Tegler, 1999; Davies et al., 2000; Barucci et al., 2000; Jewitt & Luu, 2001; Boehnhardt et al., 2001; Sheppard & Jewitt, 2002, 2003). Its popularity is due in part to its large size, and in part to its orbit. With absolute R-band magnitude $H_R$ = 4.99, Sila-Nunam is the intrinsically brightest member of the cold-classical sub-population of the Kuiper Belt (Hainaut et al., 2012). These bodies, with low inclinations, low eccentricities, and with stable non-resonant orbits beyond the 2:3 mean-motion resonance with Neptune are of special interest because they likely formed insitu (Kenyon et al., 2008; Batygin et al., 2011; Wolff et al., 2012). Having remained at their current large heliocentric distances, they provide a relatively unaltered record of the population of early solar-system bodies. The cold classicals are also distinguished by their predominantly red colors and relatively high fraction of binaries (Doressoundiram et al., 2008; Noll et al., 2008a, 2008b). With V-R = 0.67 +/- 0.05, Sila-Nunam has an optical color close to the mean for the cold classicals. Grundy et al. (2005) also report a featureless near-IR spectrum, typical of many KBOs. Owing to its low-inclination and large distance from the Sun, Sila-Nunam has been observed at extremely low phase angles (<0.01 deg), allowing probes of coherent backscatter (Rabinowitz et al., 2009; Verbiscer et al., 2010).

Recently, Sila-Nunam has become a target of heightened interest because the binary components (Sila and Nunam) are mutually eclipsing, with one body hiding the other over an 8-h interval every half orbital period, or roughly 6.25 days (Grundy et al., 2012). Such eclipses have been observed for only three other KBO systems: Pluto/Charon (Buie et al., 1992), the Haumea triple system (Fabrycky et al. 2008) and the near-contact binary (139775) 2001 QG298 (Sheppard & Jewitt, 2004). As with other binaries, the eclipses present a unique opportunity to accurately measure the size and density of the two components, and to detect small-scale changes in the color or albedo of the eclipsed body as it surface is gradually hidden and then exposed. Benecchi et al (2013) report the results of a coordinated campaign to observe one such eclipse and better constrain the binary orbit parameters. While the deepest, most complete eclipses are occurring only for the next few years, partial eclipses are expected until the end of 2017. In the proceeding discussion, we refer to Grundy et al. (2012) as Paper I and Benecchi et al. (2013) as Paper II. For eclipse ephemerides, see paper II and http://www2.lowell.edu/~grundy/tnbs/79360_1997_CS29_Sila-Nunam.html.

In this paper, we present the first precise measurement of the rotational light curve of the Sila-Nunam system, observed in several different optical band passes. While the orbital period of the binary, $P_{orb}$ = 12.510061 ± 0.000018 d, is well determined from the HST observations and the subsequent eclipse observations reported in papers I and II, the rotational state of the system has not previously been well determined. Because the binary is a tight circular system (separation 2772 ± 14 km, eccentricity 0.026 ± 0.006) it is expected to be tidally locked, similar to the Pluto/Charon system (see analysis in Paper I). Paper I presents a low signal-to-noise light curve, assembled from scattered observations taken by various telescopes over a one-year period. This tentatively supports a light curve periodicity at half the binary's orbital period, $P_{orb}/2$. Such a rotation is expected if Sila and Nunam are slightly elongated spheroids, with their long axes naturally pointed towards each other. In order to properly interpret the eclipse observations now being obtained, however, it is important to have a better measurement of the rotational light curve. This rotational light curve constrains the shapes of Sila and Nunam, sets upper limits to any variation in their color or albedo with rotational phase, and confirms that they are tidally locked.

## 2. Observations

We obtained some of the observations presented here in classical observing mode with the 2.5-m du Pont telescope of the Carnegie Institution for Science at Las Campanas, Chile. The remaining observations were obtained in service mode with the Gemini North 8.1-m telescope at Mauna Kea and the SMARTS 1.3-m at Cerro Tololo, Chile. See Tables 1 and 2 summarizing the observing circumstances and instrument characteristics. The du Pont observations were acquired in two runs, the first occurring 2011 Mar 9 to 13 and the second a year later form 2012 Mar 18 to 22. Each run was a series of six to eight 300-s exposures taken nightly with SITe2K camera, with the Gunn r filter used for all run 1, and Bessel R for all of run 2. The conditions were photometric for both runs. The Gemini and SMARTS observations were obtained in mixed conditions (photometric and non-photometric with some clouds). The Gemini observations consist of a series of five images (two 300-s exposures in Gunn g followed by three 200-s exposures in Gunn i) taken with the GMOS-N camera on each of 15 dark nights from 2012 Dec 12 to 2013 Jan 15. On a few of these nights, the imaging sequence is incomplete because of weather interruptions. The SMARTS observations consist of one or two Johnson R-band images (each a 600-s exposure) taken with the optical channel of the ANDICAM camera on each of 17 nights (mostly dark) from 2010 Dec 14 to 2011 Feb 6. Nightly averages of the SMARTS and du Pont observations appear previously in paper I and in Benecchi & Sheppard (2013). We present the un-averaged data here for a more detailed analysis.

We processed all of the SMARTS observations using bias frames and twilight flats obtained nightly, and made photometric measurements following the reduction procedure described in detail by Rabinowitz et al. (2007). Bright field stars present in the target images and observed on photometric nights are selected and calibrated with respect to Landolt (1992) standards also observed on those same photometric nights. The target flux in each exposure (including observations on non-photometric nights) is then calibrated by measuring the target flux relative to the calibrated field stars in the same image. To optimize signal to noise, we use a small aperture for these relative flux measurements (diameter 2.2", slightly larger than the typical seeing). We use the APHOT routines in IRAF to make these measurements. The typical, night-to-night systematic error using this method is ~1.5%. Note that the peak motion of Sila-Nunam at opposition is 3.0 arcsec h$^{-1}$. For the 10-min SMARTS exposures the resulting trailing is 0.5". This has a negligible effect (< 1%) on the flux measurements of the target given the relatively large measurement aperture.

We processed the Gemini data using bias frames and twilight flats gathered at regular intervals by the Gemini operators as a service to queue-scheduled observers. Aperture fluxes were then measured for every source in the processed target images using SExtractor (Bertin & Arnouts, 1996). For all flux measurements the aperture diameter was fixed at 0.87", comparable to the median seeing of the observations. Given the peak apparent motion of Sila-Nunam in our longest exposures (0.23"), the resulting flux loss due to target trailing is negligible (<1%). We then used the Sloan Digital Sky Survey (SDSS) on-line catalogue of stellar photometry (Adelman-McCarthy et al., 2008) to calibrate each field, including those acquired in non-photometric observations. This was possible because all target fields overlapped the SDSS survey area. The calibration was done in two passes. In pass 1, we determined both a zero point and color correction (proportional to g-i) separately for each target field. In pass 2, we then re-determined the zero points assuming the same g- and i-band color coefficients for all images, fixed at their median values from pass 1. By taking the same color coefficients for all fields, irrespective of the extinction from air mass or clouds, we implicitly assume that the color-dependence of the extinction does not vary with observing conditions.

For the du Pont data, we used aperture-corrected photometry to measure the point source magnitudes. The Gunn r-band and Bessel R-band data were then calibrated using photometric standards with KBO-like colors in the Gunn and Johnson systems, respectively. Details of the calibration are described by Benecchi & Sheppard (2013). To merge the two data sets into a single set useful for light curve analysis, we transformed the du Pont R-band data to the Gunn r-band system using the following transformation equation,

$$r - R = A [V-R] + B, \qquad (1)$$

where $A = 0.275 \pm 0.006$ and $B = 0.086 \pm 0.004$, appropriate for population I stars with V-R < 0.93 (Jordi et al. 2006). Assuming $V-R = 0.67 \pm 0.036$, as reported for Sila-Nunam by Hainaut et al (2012), we thereby obtain the conversion offset $r - R = 0.270 \pm 0.011$. Note that Benecchi & Sheppard (2013) perform a similar conversion of the same observations. However, they use a conversion offset, $r - R = 0.202$ mag, which did not account for Sila-Nunam's red color. The new r-band magnitudes we report here supersede the previously reported values.

Table 3 lists the resulting brightness measurements (M') and their measurement error ($\sigma_M$) for all the useful observations from the du Pont, SMARTS, and Gemini telescopes. We have excluded target measurements obtained in very poor conditions (bright moon light, poor seeing, heavy cloud cover) or contaminated by cosmic ray hits or nearby bright stars. Table 3 also lists the Julian Date (JD) at the mid-time of each exposure, the solar phase angle ($\alpha$), the target distance to the Sun ($r_s$) and to the Earth ($r_e$), the reduced magnitude ($M = M' - 5\log[r_s r_e]$), the light travel time ($\Delta t$) relative to a standard reference location and time (taken to be the first entry in the table), and the band pass of the observations. Band pass "$r_B$" is listed for those du Pont R-band data that have been transformed from Bessel R to Gunn r. In the proceeding analysis of the light curve, all observations times have been adjusted by $\Delta t$ to account for the relative motion of the target and Earth over the long time span of the observations. The largest time correction is ~5 minutes. Also, those observations occurring during predicted eclipse windows (marked with an asterisk after the band pass identifier) are not included in the analysis of the rotational light curve. Note there are no air mass corrections as the measurements are all referred to standards observed in the same field as the target. We also assume that the extinction from clouds is grey (see Rabinowitz et al., 2007).

## 3. Analysis

### 3.1 *Analysis Strategy*

When analyzing the rotational light curve of a distant body such as Sila-Nunam, observed sporadically over a time interval much larger than the rotational period, P, it is important to remove the long-term brightness variations caused by the opposition effect. With solar phase angle $\alpha$ typically ranging from 0.1 to 2.0 deg for KBOs, the resulting variation in the reduced magnitude is usually proportional to $\alpha$ with slope $\beta$ in the range 0.0 to 0.2 mag deg$^{-1}$ (Sheppard & Jewitt, 2002; Rabinowitz et al., 2007). A linear fit yields a reliable measure of $\beta$ as long as the rotational variation is insignificant, or there are enough observations for the rotational variations to average to zero. After removing the $\alpha$-dependence, an analysis of the time variability of the residual magnitude variations yields P and the shape of the rotational light curve (Rabinowitz et al. 2007, 2013).

Recognizing that Sila-Nunam's rotation period is likely to be very long, and that our observations do not uniformly cover the rotation period in all band passes, we opt in this paper to solve simultaneously for β and P. We begin by making the following simplifying assumptions:

(1) the α dependence is the same across all wavelengths.

(2) the shape of the rotational light curve is close to sinusoidal.

(3) the amplitude and phase of the light curve are the same across all wavelengths.

Previous surveys of the solar phase curves of KBOs show that assumption (1) is generally valid, but there are a few exceptions (Rabinowitz et al., 2007). Assumption (2) is generally not a valid assumption. When observed with high precision, some small solar-system bodies do not have sinusoidal light curves. Often there are two peaks per rotation cycle with dissimilar amplitudes and widths resulting from the irregular shape of the body (Sheppard & Jewitt, 2002; Sheppard, 2007; Benecchi & Sheppard, 2013). In the case of the Pluto-Charon system, the light curve shape is non-sinusoidal because the rotational variation is related to albedo patterns, not the shape of the bodies (Tholen & Tedesco, 1994). Nonetheless, assumption (2) allows us to fit for a fundamental period, which would be half the true rotation period for a double-peaked light curve. After finding the fundamental period, we then relax this assumption and examine the evidence for a double-peaked light curve (see Sec. 4.2). Assumption (3) is usually valid for most KBOs because it is the rotational variation of their projected area that determines their light curve shape. Wavelength-dependent rotational light curve shapes are rare, but there are a few noteworthy cases such as Pluto (Grundy & Buie, 2001), 136108 Haumea (Lacerda, 2009), and more recently 2010 WG$_9$ (Rabinowitz et al., 2013)

Given the above assumptions, we model the time dependence of the reduced magnitude in each band pass, j, by the following expression:

$$M(t) = H_j + \beta\, \alpha(t) + F(t) \qquad (2)$$

where $H_j$ is the mean reduced magnitude extrapolated to α = 0 (i.e. the absolute magnitude in band j) and F(t) describes the rotational light curve. We consider two separate possibilities: $F(t) = F_s(t)$ and $F_d(t)$, where $F_s(t)$ is a simple, single-peaked sinusoidal function and $F_d(t)$ is a more complex double-peaked sinusoid (see below). As mentioned above, we first take $F(t) = F_s(t)$ to determine a fundamental period. Holding the period at this value, we then take $F(t) = F_d(t)$ to fit more complex features.

We parameterize $F_s(t)$ as follows:

$$F_s(t) = A\sin(2\pi\,[\phi + \omega(P,t)]) \qquad (3)$$

Here $\omega(P,t)$ is the rotational phase, $\omega(P,t)=(t-t_o)/P$, at time, t. Parameters A and $\phi$ are the fixed harmonic amplitude and phase offset, and $t_o$ is an arbitrary fixed epoch. A full description of the light curve in the 4 observed bands requires 8 free parameters ($H_1$, $H_2$, $H_3$, $H_4$, $\beta$, A, $\phi$, and P).

We express $F_d(t)$ as the sum of two sinusoidal terms:

$$F_d(t) = A_o \sin(2\pi [\phi_o + 2\omega(P,t)]) + A_\Delta \sin(2\pi[\phi_\Delta + \omega(P,t)] ) \qquad (4)$$

The first term describes what we expect to be the dominant signal, a simple sinusoid with two peaks per rotation period. The second term describes a small, perturbing sinusoid with one peak per rotation cycle. The phase and amplitude of the two terms are independent, but they share the same period. A full description of the light curve requires 10 free parameters ($H_1$, $H_2$, $H_3$, $H_4$, $\beta$, P, $A_o$, $\phi_o$, $A_\Delta$, $\phi_\Delta$).

The above form for $F_d(t)$ describes the variability we would expect if Sila and/or Nunam not only have an elliptical shape, but also have non-uniform albedo distributions. For example, a relatively bright or dark hemisphere would modulate the apparent brightness only once per rotation, with this variation superimposed on the double-peaked light curve due to ellipsoidal shape. Such an unusual albedo distribution might be expected from an uneven distribution of surface volatiles, an ancient impact crater exposing sub-surface material with optical properties different from the surface, or from surface alteration caused by exchange of impact ejecta between Sila and Nunam (Stern, 2009).

For either of our two formulations for the rotational light curve, we solve for the respective parameters by minimizing the following expression:

$$\chi^2 = \Sigma_j \chi_j^2 \qquad (5)$$

where for each band pass

$$\chi_j^2 = \Sigma_i [M_i - H_j - \beta \times \alpha_i - F(t_i)]^2 / \sigma_i^2 \qquad (6)$$

The sum is over each observation, i, with corresponding magnitude $M_i$, observation time, $t_i$, solar phase angle, $\alpha_i$, and measurement uncertainty, $\sigma_i$. Note that the absolute magnitudes, $H_j$, are independently constrained once the parameters defining F(t) are fixed. For each band pass, they are the weighted average value of the reduced magnitude after $\alpha$ and rotation dependence are subtracted:

$$H_j = [\Sigma_i (1/\sigma_i^2)]^{-1} \Sigma_i [M_i - \beta \times \alpha_i - F(t_i)]/\sigma_i^2 \qquad (7)$$

3.2 *Data Selection and Averaging*

   Before proceeding with the above analysis, we implement the following procedure to reduce the scatter in the observations and to cull outliers. The procedure is implemented separately for the observations in each band pass.

1. We exclude observations observed at solar phase angle $\alpha < 0.1$ deg. In this range, some bodies exhibit an opposition spike departing strongly from their linear behavior at larger phase angles (Verbiscer et al., 2007; Buratti et al., 2011). We reserve a more detailed discussion of these low-phase angle observations for a future paper.

2. We exclude observations obtained during the 8-hour window of any observed or predicted eclipse. Such observations would clearly deviate significantly from the simple light curve shape we are assuming. In Sec 4.5, we re-examine these observations after we have modeled the un-eclipsed brightness modulations.

3. We require at least two observations per night and take the weighted average magnitude for each night. Because we are searching for a s srotation period comparable to the half-orbit period of 6.25 days, the nightly averages allow us to increase the signal to noise of our data samples without significantly affecting the analysis of the rotational variability. Note that all the observations we consider for light curve fitting are obtained within a 15-min interval on any given night.

4. After nightly averaging, we add a systematic error in quadrature to the resulting measurement uncertainties to account for small, night-to-night variations in the flux calibrations. Such variations are the unavoidable outcome of targeting moving objects, for which a common set of field stars cannot be used to calibrate all the observations. There are also possible measurement errors as the target moves close to faint sources at the detection limit of the images, with these sources changing from night to night. Previous observations with the SMARTS telescope show that an 0.015-mag variance is realistic (Rabinowitz et al., 2007).

5. We exclude nightly averaged observations with uncertainty > 0.15 mag.

6. After applying the above procedure, we identify outliers by subtracting an initial linear fit to the $\alpha$-dependence, calculating the mean residual, and excluding those observations with residual magnitudes differing by more than 0.2 mag from the mean. This variance threshold is nearly three times the rotational variance determined for Sila-Nunam in Paper I ($\pm$ 0.07 mag).

7. We exclude the du Pont observations from the nights of 2011 Mar 12 and 2012 Mar 17 ( JD-2445000 = 5633 and 6004). For unknown reasons, these observations are anomalously bright by several tenths of a mag, which is much larger than their typical uncertainty (~0.02 mag after nightly averaging). This may be the result of confusion with a faint source or calibration error.

Note that steps 1-6 have no effect on the Gemini and du Pont observations other than to yield nightly averages. The SMARTS data are significantly restricted, however, because their measurement uncertainty is large and some of the observations were obtained during eclipses. Note that restriction in $\alpha$ (step 1) has little bearing on the results of this paper since only a few SMARTS observations are removed.

## 4.0 Results and Discussion

### 4.1 Best-Fit Light Curve Parameters for a Single-Peaked Sinusoid

To find the best-fit solution for M(t) assuming single-peaked sinusoidal variation, $F(t) = F_s(t)$, we minimized Eq. (5) using a brute-force evaluation, keeping track of the minimum $\chi^2$ while stepping all 8 free parameters through wide search ranges. With fixed $t_o = 2455500$, we varied A from 0.0 to 0.2 mag by steps of 0.01 mag, $\beta$ from 0.0 to 0.3 mag deg$^{-1}$ by 0.01 mag deg$^{-1}$, $\phi$ over its full range(0 to 1) by 0.01, and P from 1 to 20 d in 10000 uniform logarithmic increments. For each set of fixed values for A, $\beta$, $\phi$, and P, the values for the remaining free parameters ($H_j$ with j = 1 to 4) were determined using Eq. (7). Near the values of A, $\beta$, and P at the resulting minimum, we then repeated the evaluation with roughly double the resolution in the search grid (still stepping $\phi$ over its full search range). Table 4 (first row) lists the resulting best-fit values of P, $\beta$, A, and $\phi$. Also listed are the respective uncertainties in each parameter ($\sigma_P$, $\sigma_\beta$, $\sigma_A$, and $\sigma_\phi$), the minimum $\chi^2$, the number of degrees of freedom ($\nu$), and the chi-square likelihood (L). The uncertainties were determined by finding the range of values for P, $\beta$, A, and $\phi$ for which $\chi^2$ remained less than the minimum value plus one. These are the 68.3% confidence interval for each parameter evaluated individually (Press et al. 1986).

We see from Table 4 that the best-fit sinusoid has minimum $\chi^2$ = 45.98 and L = 0.052 at P = 6.2562 ± 0.002 d with $\nu$ = 32 degrees of freedom. The small likelihood for this fit may indicate that a simple single-peaked sinusoid is not appropriate, or that we have underestimated the measurement uncertainties for some of the observations. Nonetheless, the resulting $\chi^2$ is much less than value we obtain with no sinusoidal correction ($\chi^2$ = 127.7), which is the minimum value we find keeping A = $\phi$ = P = 0. This large reduction in $\chi^2$ is a strong indication that we have measured a significant periodicity. Also, the best-fit period differs by only 0.02 % from the half orbital period ($P_{orb}/2$ = 6.25503 +/- 0.00009 d) measured independently in Papers I and II. Thus it is almost certain that we have detected synchronous rotation. Note that we obtain consistent solutions fitting the Gemini g and i band data alone, but with larger error bars for the resulting rotational period.

To show the uniqueness of our solution for P, we present a periodogram for Sila-Nunam in Figure 1. This is a plot of the minimum $\chi^2$ we find versus P for each of the fixed values of P that we evaluated in our search for the overall best fit. At each value, we varied the remaining degrees of freedom (A, $\beta$, and $\phi$) over their full search ranges. Panel (a) shows the complete periodogram, while (b) shows an expanded view near the deepest minimum. Dashed horizontal lines indicate the formal 68.3% and 95.4% confidence limits, and the dashed vertical line marks the value of $P_{orb}/2$.

It is apparent that there are secondary local minima on each side of the best-fit period, with the next-lowest minima occurring at $\chi^2 = 53.3$ (with formal likelihood 0.01). They occur at aliases of the true period, resulting from the long time interval (~600 days) between the earlier du Pont and SMARTS observations and the later Gemini observations. To measure the likelihood that the chi-square minimum occurs at the true period, we performed a Monte Carlo analysis of our observations. Repeatedly fitting simulated observations of our best-fit sinusoidal light-curve, observed at the same times and with the same measurement errors as the real data, we find that the chi-square minimum coincides with the true period 89% of the time. We also confirm that our 1-sigma measurement uncertainty for the true period is ~0.002d. Hence, we have 89% confidence we have measured the correct period and that it is consistent with $P_{orb}/2$ at the 1-sigma level of uncertainty stated above.

With a unique solution for P occurring very close to $P_{orb}/2$, we can safely assume that $P_{orb}$ is the true rotation period. The reason we observe a light curve period with half this value is that Sila or Nunam or both bodies has an ellipsoidal shape. The small deviation from $P_{orb}/2$ that we measure is very likely the result of measurement error alone. It would otherwise be physically unlikely to find the rotation very close, but not locked to the orbital motion. A small deviation would imply a relatively recent collisional event (within the last 0.1 to 1 Gyr) that either formed the binary or excited the rotation of Sila or Nunam (see discussion in Paper I). Such collisional events are very unlikely (Levison et al., 2008). If we assume that Sila-Nunam is indeed synchronously locked and fix $P = P_{orb}/2$, then our best-fit values for A and β and their uncertainties remain unchanged (see second row of Table 4). Parameter ϕ changes marginally because it is tightly correlated with P, while its uncertainty drops by a factor ~3.

*4.2 Best-Fit Light Curve Parameters for a Double-Peaked Sinusoid*

Having established that the rotation period is synchronous, we can now refit the observations assuming the more complex rotational variation, $F(t) = F_d(t)$, consisting of summed double- and single-peaked sinusoids. Keeping the period fixed at $P = P_{orb}$, we thus obtained new minimum $\chi^2 = 33.30$. Table 5 lists the corresponding best-fit values for the double peaked amplitude and phase ($A_o$, $\phi_o$), the single peaked amplitude and phase ($A_\Delta$, and $\phi_\Delta$), and β. Figure 2 shows the observed, phase-folded rotational light curve in each band pass, which is the α-corrected reduced magnitude, $\{M_i - \beta\alpha_i\}$, as a function of rotational phase, $\omega(P, t_i)$, taking $P = P_{orb}$. Superimposed on each light curve is the best-fit, double-peaked sinusoid, $H_j + F_d(t)$. Note that we have arbitrarily shifted each light curve vertically for clarity. We have also redefined zero rotational phase to coincide with the minimum for $F_d(t)$. Also note that all the SMARTS and du Pont observations acquired near zero phase ($\phi = -0.03$ to $0.03$) occur during mutual events. These observations are excluded from the determination of $F_d(t)$.

The resulting solution for $F_d(t)$ is dominated by the double-peaked term with harmonic amplitude $A_o = 0.060 \pm 0.006$ mag and phase $\phi_o = 0.265 \pm 0.015$. This is essentially unchanged from the best-fit, single-peaked sinusoid, $F_s(t)$, we obtained earlier assuming $P = P_{orb}/2$ (for which $A = 0.055 \pm 0.005$ mag and $\phi = 0.287 \pm 0.015$). The new solar-phase coefficient, $\beta =$ 0.147 ± 0.018 mag deg$^{-1}$, is also nearly the same. However, the new single-peaked component is a significant alteration having non-negligible harmonic amplitude, $A_\Delta = 0.022 \pm 0.005$ mag.

Inspecting Fig. 2, it is clear that the double-peaked sinusoid is a good fit to the observations. It is also a significantly better fit than the solution we obtain assuming only a simple single-peaked sinusoid, $F_s(t)$. The resulting reduction in $\chi^2$ (from 46.35 to 33.30) exceeds the chance reduction we would expect simply because we have introduced two new parameters and thereby decreased the number of fitted degrees of freedom (from $\nu = 33$ to 31). Using the F-test (Bevington, 1992), we calculate statistic F = [difference in $\chi^2$/ difference in $\nu$] / [$\chi^2/\nu$] = 6.07. This has likelihood 0.6% under the null hypothesis that the new parameters do not significantly reduce $\chi^2$. Hence we have 99.4% confidence that $F_d(t)$ is a better match to the observations than $F_s(t)$. We also ran a Monte Carlo program to calculate the likelihood of obtaining this solution by chance if the true light curve were described by $F_s(t)$. Repeatedly simulating observations at the same times and with the same uncertainty as the real observations, we find a likelihood of 1.8% for obtaining a solution with $A_\Delta > 0.02$ mag. Hence, we have 98% confidence that the single-peaked feature is not the result of measurement error.

The above solution for $F_d(t)$ also provides a good fit for the observations in each band pass taken individually. Table 5 lists the respective $\chi^2$ values we obtain using Eq. (7) to solve for each value of H. The table also lists the respective the number of fitted observations, N, the uncertainty, $\sigma_H$, and the chi-square likelihood, L. Note that for each band there is only one fitted parameter, H, and the corresponding number of degrees of freedom is N-1. In all cases, we obtain $\chi^2 \sim$ N-1 and L > 10%, values we would expect for a good fit. We note, however, that we obtain $H_r - H_R = 0.196 \pm 0.032$. This is inconsistent with the value r-R = 0.270 ± 0.011 we assume to merge the du Pont observations in the r and R bands (see Sec 2). We believe the later value is correct, however, because the resulting scatter of the merged du Pont observations is ~0.01 mag with respect to our best-fit light curve. Instead, it is possible that there is a systematic error in our calibration of the SMARTS R-band observations of ~0.07 mag, perhaps relating to the extreme red color of Sila-Nunam or to the small trailing of the target during the long SMARTS integrations. The error would not affect any of the conclusions of this paper.

4.3 *Color Dependence to the Rotational Light Curve*

To explore the possibility of a small wavelength dependence to the rotational light curve, we replot the rotationally phased observations in Figure 3a. These are the same data represented in Fig. 2 (nightly averaged reduced magnitudes corrected for solar phase dependence). Here, however, we subtract the best-fit values for H in each band (see Table 6). This shifts all the light curves to the same mean. Under our assumption that there is no color dependence to the rotational modulation, all the observations should combine to yield a consistent light curve. Figure 3a also shows our best-fit double-peaked light curve, $F_d(t)$, while Figure 3b shows the residuals after subtracting $F_d(t)$ from the observations.

If there were any wavelength-dependence to the light curve, the signal would appear in Figure 3b as a non-random dependence upon rotational phase in a particular band pass. Inspection of the figure shows that there is no significant evidence for such a pattern. There is only a suggestion of a dependence in the Gemini g band between rotational phases -0.37 and -0.027 where the residuals (from 2 separate nights) are positive by 0.02 ± 0.02 and 0.04 ± 0.02 mag, respectively. Since this is the largest such feature, it sets an upper limit of ~0.04 mag for the possible variation in the g-i color. Additional observations would be required to determine its validity.

### 4.4 *Eclipse Observations*

Examining the light curves for SMARTS R and du Pont r in Figure 2, we see that the eclipse observations are centered very closely to zero phase, where the best-fit double-peaked sinusoid with $P = P_{orb}$ reaches a minimum. The zero-phase coincidence is further evidence that the rotation of Sila and/or Nunam is synchronously locked to the orbit. If either body is elongated, tidal forces will bring its axis of elongation into alignment with the line separating the two bodies. We would then expect a minimum in the rotational light curve when we are viewing the smallest projected area of the rotating body, which would necessarily coincide with the eclipses.

Figure 4 provides an expanded view of these eclipse data. Also plotted are model predictions for the eclipse light curves for the two events best covered by the observations (dates JD 2455593 and 2456006). These predictions are based on the binary orbital parameters reported in papers I and II and assume that both Sila and Nunam are Lambertian spheres. Inspecting the figure, we see that the eclipse observations are consistent with the model predictions. Given the measurement uncertainties and the incomplete coverage of the mutual events, these data are not useful for further constraining the orbital model nor the relative sizes of Sila and Nunam. Of more significance, the figure shows that the predicted eclipse minima are coincident with zero rotational phase to within 0.3% of a rotation cycle. This shows that the long axis of Sila and/or Nunam is precisely aligned with the other body.

### 4.5 *Shape Constraints*

We can constrain the shapes of Sila and Nunam from the amplitude of their mutual light curve assuming they are both (a) synchronously rotating, (b) near in shape to triaxial (Jacobian) ellipsoids, and (c) rotating about their shortest axis, with their spin vectors aligned with their mutual orbital angular momentum vector and their long axes pointed toward the mutual center of mass. Assumption (a) is likely because the two bodies are nearly the same size (diameter ~240 km, see Paper I). Tidal forces synchronizing their separate rotations are therefore roughly equivalent, and should affect both bodies similarly. Assumption (b) is likely because self gravity for bodies larger than ~200 km is strong enough to overcome modest internal strengths and force their figures into quasi-hydrodynamic equilibrium. This is evidenced by resolved images of large main-belt asteroids (Marchis et al., 2006) and by the relatively low-amplitude light curves observed for most KBOs larger than ~200 km (Benecchi & Sheppard, 2013). Assumption (c) is likely because it is the natural end state for a tidally evolved binary (see Goldreich and Peale, 1970 and Cheng et al., 2014). Also, our observation that the eclipses occur at light curve minima requires that at least one of the two bodies satisfies the assumed geometry. This geometry also implies that the spin axes of the two bodies are perpendicular to the line of sight, since we are viewing the binary orbit edge on during mutual events.

Given the above assumptions, and given a measured rotational light curve with peak-to-peak amplitude 2A magnitudes, we then have

$$(\rho_1 a_1^2 + \rho_2 a_2^2)/(\rho_1 b_1^2 + \rho_2 b_2^2) = 10^{0.8A} \qquad (10)$$

where $a_1$ and $a_2$ are the long axis dimensions, $b_1$ and $b_2$ are the intermediate axis dimensions, and $\rho_1$ and $\rho_2$ are the albedos of the two bodies. From the HST and Spitzer observations reported in paper I, we also know that Sila-Nunam are of nearly equal average brightness, $(\rho_1 a_1 b_1)/(\rho_2 a_2 b_2) = 1.12$. To better than 1% precision, we can then approximate Eq. (10) by

$$(a_1/b_1)(a_2/b_2) = 10^{0.8A} \qquad (11)$$

The above expression restricts a/b to the range $10^{0.4A}$ to $10^{0.8A}$, where the lower limit applies when both bodies have the same shape and the upper limit applies when one of two is spherical. With our measurement A = 0.060 ± 0.005, we thus find $a_1/b_1$ = 1.0 to 1.06 for one body, and $a_2/b_2$ = 1.12 $(a_1/b_1)^{-1}$ = 1.06 to 1.12 for the other. Note that if we relax the assumption that the spins of Sila and Nunam are aligned with their orbital angular momentum vector, then these values are lower bounds for their a/b ratio.

The minimum axis ratios we estimate for at least one of the bodies is larger than expected for synchronously-rotating strengthless bodies in hydrodynamic equilibrium. For example, Descamps (2010) computes equilibrium shapes for equal-mass binaries with a range of porosities. The range of orbital parameters (rotation period, separation) considered by the author do not overlap the case of Sila-Nunam. However, even for the case of equal mass binaries 1.5 times closer than Sila and Nunam, and with rotational velocities 2 times larger, the expected value for a/b is 1.02 or smaller. From this we can conclude that Sila and/or Nunam have at least some internal strength. A more extensive calculation would be required to establish limiting values for this parameter.

**5.0 Conclusions**

We have combined long-term, multi-band, photometric observations of Sila-Nunam's light curve obtained with small, medium, and large aperture telescopes to obtain a precise measurement of the rotational light curve and solar phase coefficient. We find convincing evidence that the rotation of at least one body is synchronized to the binary orbit, with a rotation period matching the orbit period previously measured with HST direct imaging and ground-based eclipse observations. After subtracting a linear solar phase dependence and ignoring observations phase angle, we are able to fit a rotational light curve with a simple sinusoidal function. However, we find a significantly better match with a double-peaked function consisting of a primary sinusoidal function with two cycles per orbit modulated by an additional, smaller-amplitude sinusoid with one cycle per orbit. This indicates that one or both bodies are elongated, and there is a unique feature (perhaps an albedo spot or crater) on one or both bodies appearing only once per rotation. We also observe that eclipses are centered at light curve minima to within 0.3%, requiring the long axis of at least one of the two bodies to point precisely toward the other. Depending in detail on the geometry of the binary orbit, the amplitude we measure for the rotational light curve jointly constrains triaxial shape models for both bodies such that the product of their long-to-intermediate axis ratios is 1.12 ± 0.01. This means that both bodies are elongated by 6%, or else is one is elongated by 6 to 12 %, and the other by less than 6%. Our observations rule out any rotational color variation at optical wavelengths exceeding ~4%.

With the conclusion that Sila-Nunam binary is synchronous, and very likely doubly synchronous, ongoing and future observations of the Sila-Nunam mutual events can now be precisely modeled to determine the shapes and relative size of the two bodies and to map color and albedo variations across their surfaces. Such studies will provide the first detailed surface characterization of an ultra-red, distant body that likely formed beyond the present orbit of Neptune. This work will be fundamental to understanding the physical composition and formation of the Kuiper Belt.

This work was supported by NASA grants NNX10AB31G and NNX09AC99G and NSF Planetary Astronomy grant AST-1109872. Special thanks go to the Gemini observers and support scientists, especially J. Rhee, M. Bergmann, A. Adamson, and A. Nitta. We also thank A. Harris and W. Fraser for detailed and constructive reviews.

**Table 1. Observing Circumstances**

| Telescope | Instrument | Filter | Exp. Time (s) | No. Nights | Start and End Date | Median Seeing (arcsec) | Conditions |
|---|---|---|---|---|---|---|---|
| SMARTS 1.3m | ANDICAM | R | 600 | 17 | 2010 Dec 14 2011 Feb 6 | 1.5 | mixed |
| du Pont 2.5m | SITe2K | r | 300 | 5 | 2011 Mar 9 2011 Mar 13 | 1.7 | photometric |
| du Pont 2.5m | SITe2K | R | 300 | 5 | 2012 Mar 18 2012 Mar 22 | 1.6 | photometric |
| Gemini North | GMOS-N | g, i | 300, 200 | 15 | 2012 Dec 12 2013 Jan 15 | 0.9 | mixed |

**Table 2. Instrument Characteristics**

| Instrument | Pixel Scale (") | Array Dim. | Binning |
|---|---|---|---|
| ANDICAM | 0.37 | 1024 x 1024 | 2 x 2 |
| SITe2K | 0.26 | 2048 x 2048 | 1 x 1 |
| GMOS-N | 0.15 | 3072 x 2304 | 2 x 2 |

Table 3. **Photometric Observations of (79360) Sila-Nunam**

| JD-2450000 | M(mag) | $\sigma_M$(mag) | $\alpha$(deg) | $r_s$(AU) | $r_e$(AU) | $\Delta t$(min) | M'(mag) | Filter | Telescope |
|---|---|---|---|---|---|---|---|---|---|
| 3761.59694 | 4.236 | 0.126 | 0.012 | 43.543 | 42.558 | 0.00000 | 20.575 | I | SMARTS |
| 3761.60460 | 4.696 | 0.093 | 0.012 | 43.543 | 42.558 | 0.00000 | 21.035 | R | SMARTS |
| 3762.60431 | 4.069 | 0.134 | 0.030 | 43.543 | 42.558 | 0.00144 | 20.408 | I | SMARTS |
| 3762.61198 | 4.641 | 0.098 | 0.030 | 43.543 | 42.558 | 0.00144 | 20.980 | R | SMARTS |
| 3772.73196 | 4.376 | 0.121 | 0.265 | 43.542 | 42.577 | 0.16272 | 20.716 | I | SMARTS |
| 3772.73962 | 5.031 | 0.096 | 0.265 | 43.542 | 42.578 | 0.16272 | 21.371 | R | SMARTS |
| 3793.62411 | 4.341 | 0.120 | 0.714 | 43.542 | 42.713 | 1.28736 | 20.688 | I | SMARTS |
| 3793.63230 | 4.974 | 0.110 | 0.715 | 43.542 | 42.713 | 1.28880 | 21.321 | R | SMARTS |
| 3794.61983 | 4.369 | 0.263 | 0.734 | 43.542 | 42.722 | 1.36656 | 20.717 | I | SMARTS |
| 3794.62749 | 4.689 | 0.204 | 0.734 | 43.542 | 42.722 | 1.36656 | 21.037 | R | SMARTS |
| 5545.81454 | 5.018 | 0.138 | 0.989 | 43.503 | 42.856 | 2.47824 | 21.371 | R | SMARTS |
| 5545.85164 | 4.923 | 0.163 | 0.989 | 43.503 | 42.855 | 2.47392 | 21.276 | R | SMARTS |
| 5547.76312 | 4.750 | 0.126 | 0.959 | 43.503 | 42.831 | 2.26800 | 21.101 | R | SMARTS |
| 5547.79282 | 4.783 | 0.099 | 0.959 | 43.503 | 42.830 | 2.26368 | 21.134 | R | SMARTS |
| 5557.80597 | 4.935 | 0.199 | 0.786 | 43.502 | 42.714 | 1.29312 | 21.280 | R | SMARTS |
| 5557.85908 | 5.013 | 0.265 | 0.785 | 43.502 | 42.713 | 1.28880 | 21.358 | R | SMARTS |
| 5561.74549 | 5.098 | 0.130 | 0.711 | 43.502 | 42.674 | 0.96624 | 21.441 | R | SMARTS |
| 5561.79498 | 5.203 | 0.130 | 0.710 | 43.502 | 42.674 | 0.96192 | 21.546 | R | SMARTS |
| 5563.78363 | 4.808 | 0.119 | 0.670 | 43.502 | 42.655 | 0.80928 | 21.150 | R | SMARTS |
| 5563.83207 | 5.119 | 0.163 | 0.669 | 43.502 | 42.655 | 0.80640 | 21.461 | R | SMARTS |
| 5564.73513 | 5.083 | 0.145 | 0.651 | 43.502 | 42.647 | 0.74016 | 21.425 | R | SMARTS |
| 5564.79399 | 5.040 | 0.136 | 0.650 | 43.502 | 42.646 | 0.73584 | 21.382 | R | SMARTS |
| 5566.70414 | 5.229 | 0.169 | 0.611 | 43.502 | 42.630 | 0.60192 | 21.570 | R | SMARTS |
| 5566.75901 | 4.808 | 0.096 | 0.610 | 43.502 | 42.630 | 0.59760 | 21.149 | R | SMARTS |
| 5569.72969 | 5.319 | 0.147 | 0.547 | 43.502 | 42.607 | 0.40752 | 21.659 | R | SMARTS |
| 5569.79941 | 5.114 | 0.144 | 0.546 | 43.502 | 42.606 | 0.40320 | 21.454 | R | SMARTS |
| 5571.80758 | 4.896 | 0.112 | 0.503 | 43.502 | 42.592 | 0.28512 | 21.235 | R | SMARTS |
| 5576.72040 | 4.964 | 0.131 | 0.395 | 43.502 | 42.563 | 0.04032 | 21.302 | R | SMARTS |
| 5576.77766 | 4.894 | 0.085 | 0.394 | 43.502 | 42.563 | 0.03744 | 21.232 | R | SMARTS |
| 5590.68681 | 5.005 | 0.097 | 0.074 | 43.502 | 42.518 | -0.32976 | 21.340 | R | SMARTS |
| 5590.73312 | 4.795 | 0.083 | 0.073 | 43.502 | 42.518 | -0.32976 | 21.130 | R | SMARTS |
| 5592.59969 | 4.936 | 0.107 | 0.030 | 43.502 | 42.517 | -0.34128 | 21.271 | R | SMARTS |
| 5592.63137 | 4.872 | 0.090 | 0.029 | 43.502 | 42.517 | -0.34128 | 21.207 | R | SMARTS |
| 5592.66378 | 4.984 | 0.096 | 0.028 | 43.502 | 42.517 | -0.34128 | 21.319 | R | SMARTS |
| 5592.69773 | 5.013 | 0.097 | 0.027 | 43.502 | 42.517 | -0.34128 | 21.348 | R | SMARTS |
| 5592.74058 | 4.942 | 0.098 | 0.026 | 43.502 | 42.517 | -0.34128 | 21.277 | R | SMARTS |
| 5592.78068 | 4.978 | 0.103 | 0.026 | 43.502 | 42.517 | -0.34128 | 21.313 | R | SMARTS |
| 5593.62562 | 5.399 | 0.196 | 0.010 | 43.502 | 42.517 | -0.34416 | 21.734 | R* | SMARTS |
| 5593.67535 | 5.353 | 0.156 | 0.009 | 43.502 | 42.517 | -0.34416 | 21.688 | R* | SMARTS |
| 5593.70720 | 5.130 | 0.130 | 0.009 | 43.502 | 42.517 | -0.34416 | 21.465 | R* | SMARTS |
| 5593.71835 | 5.131 | 0.148 | 0.009 | 43.502 | 42.517 | -0.34416 | 21.466 | R* | SMARTS |
| 5593.74505 | 5.120 | 0.264 | 0.008 | 43.502 | 42.517 | -0.34416 | 21.455 | R* | SMARTS |
| 5593.76631 | 5.006 | 0.181 | 0.008 | 43.502 | 42.517 | -0.34416 | 21.341 | R* | SMARTS |
| 5594.62520 | 4.770 | 0.088 | 0.021 | 43.502 | 42.517 | -0.34272 | 21.105 | R | SMARTS |
| 5594.66001 | 4.912 | 0.088 | 0.021 | 43.502 | 42.517 | -0.34272 | 21.247 | R | SMARTS |
| 5594.71328 | 5.047 | 0.099 | 0.022 | 43.502 | 42.517 | -0.34272 | 21.382 | R | SMARTS |
| 5594.72498 | 4.863 | 0.090 | 0.022 | 43.502 | 42.517 | -0.34272 | 21.198 | R | SMARTS |
| 5594.74744 | 5.004 | 0.110 | 0.023 | 43.502 | 42.517 | -0.34272 | 21.339 | R | SMARTS |
| 5596.77688 | 4.752 | 0.118 | 0.069 | 43.502 | 42.518 | -0.33408 | 21.087 | R | SMARTS |
| 5597.70103 | 4.846 | 0.093 | 0.091 | 43.502 | 42.519 | -0.32688 | 21.181 | R | SMARTS |
| 5597.78893 | 4.974 | 0.148 | 0.093 | 43.502 | 42.519 | -0.32544 | 21.309 | R | SMARTS |
| 5599.68190 | 5.015 | 0.118 | 0.137 | 43.502 | 42.522 | -0.30240 | 21.351 | R | SMARTS |
| 5599.73522 | 5.084 | 0.133 | 0.138 | 43.502 | 42.522 | -0.30240 | 21.420 | R* | SMARTS |
| 5629.53536 | 5.190 | 0.042 | 0.780 | 43.501 | 42.704 | 1.21536 | 21.535 | r | du Pont |
| 5629.53963 | 5.291 | 0.042 | 0.781 | 43.501 | 42.704 | 1.21536 | 21.636 | r | du Pont |
| 5629.55700 | 5.286 | 0.042 | 0.781 | 43.501 | 42.704 | 1.21680 | 21.631 | r | du Pont |
| 5629.57458 | 5.274 | 0.042 | 0.781 | 43.501 | 42.704 | 1.21824 | 21.619 | r | du Pont |
| 5629.57885 | 5.217 | 0.042 | 0.781 | 43.501 | 42.705 | 1.21824 | 21.562 | r | du Pont |
| 5629.63373 | 5.236 | 0.042 | 0.782 | 43.501 | 42.705 | 1.22400 | 21.581 | r | du Pont |
| 5629.63800 | 5.202 | 0.042 | 0.782 | 43.501 | 42.705 | 1.22400 | 21.547 | r | du Pont |
| 5629.64228 | 5.175 | 0.042 | 0.783 | 43.501 | 42.705 | 1.22400 | 21.520 | r | du Pont |
| 5630.52333 | 5.331 | 0.039 | 0.799 | 43.501 | 42.714 | 1.29888 | 21.676 | r | du Pont |
| 5630.52760 | 5.370 | 0.039 | 0.799 | 43.501 | 42.714 | 1.30032 | 21.715 | r | du Pont |
| 5630.54506 | 5.349 | 0.039 | 0.799 | 43.501 | 42.714 | 1.30176 | 21.694 | r | du Pont |
| 5630.56236 | 5.258 | 0.039 | 0.800 | 43.501 | 42.715 | 1.30320 | 21.603 | r | du Pont |
| 5630.56663 | 5.276 | 0.039 | 0.800 | 43.501 | 42.715 | 1.30320 | 21.621 | r | du Pont |
| 5630.61762 | 5.292 | 0.039 | 0.801 | 43.501 | 42.715 | 1.30752 | 21.637 | r | du Pont |
| 5630.62190 | 5.275 | 0.039 | 0.801 | 43.501 | 42.715 | 1.30752 | 21.620 | r | du Pont |
| 5630.62617 | 5.331 | 0.039 | 0.801 | 43.501 | 42.715 | 1.30896 | 21.676 | r | du Pont |
| 5631.51887 | 5.185 | 0.056 | 0.817 | 43.501 | 42.725 | 1.38672 | 21.531 | r | du Pont |
| 5631.52314 | 5.252 | 0.056 | 0.817 | 43.501 | 42.725 | 1.38672 | 21.598 | r | du Pont |
| 5631.54042 | 5.256 | 0.056 | 0.818 | 43.501 | 42.725 | 1.38816 | 21.602 | r | du Pont |
| 5631.55779 | 5.369 | 0.056 | 0.818 | 43.501 | 42.725 | 1.38960 | 21.715 | r | du Pont |
| 5631.56206 | 5.320 | 0.056 | 0.818 | 43.501 | 42.725 | 1.39104 | 21.666 | r | du Pont |
| 5631.61818 | 5.257 | 0.056 | 0.819 | 43.501 | 42.726 | 1.39536 | 21.603 | r | du Pont |
| 5631.62245 | 5.281 | 0.056 | 0.819 | 43.501 | 42.726 | 1.39536 | 21.627 | r | du Pont |
| 5631.62672 | 5.346 | 0.056 | 0.819 | 43.501 | 42.726 | 1.39680 | 21.692 | r | du Pont |
| 5632.51654 | 5.270 | 0.054 | 0.835 | 43.501 | 42.735 | 1.47600 | 21.616 | r | du Pont |
| 5632.52953 | 5.168 | 0.054 | 0.836 | 43.501 | 42.736 | 1.47744 | 21.514 | r | du Pont |
| 5632.54259 | 5.343 | 0.054 | 0.836 | 43.501 | 42.736 | 1.47888 | 21.689 | r | du Pont |
| 5632.56779 | 5.211 | 0.054 | 0.836 | 43.501 | 42.736 | 1.48032 | 21.557 | r | du Pont |
| 5632.57207 | 5.276 | 0.054 | 0.836 | 43.501 | 42.736 | 1.48032 | 21.622 | r | du Pont |
| 5632.57635 | 5.240 | 0.054 | 0.836 | 43.501 | 42.736 | 1.48176 | 21.586 | r | du Pont |
| 5633.51793 | 5.110 | 0.052 | 0.853 | 43.501 | 42.746 | 1.56672 | 21.457 | r | du Pont |
| 5633.52220 | 5.103 | 0.052 | 0.853 | 43.501 | 42.747 | 1.56816 | 21.450 | r | du Pont |
| 5633.53948 | 5.169 | 0.052 | 0.854 | 43.501 | 42.747 | 1.56960 | 21.516 | r | du Pont |
| 5633.55682 | 5.212 | 0.052 | 0.854 | 43.501 | 42.747 | 1.57104 | 21.559 | r | du Pont |
| 5633.56109 | 5.153 | 0.052 | 0.854 | 43.501 | 42.747 | 1.57104 | 21.500 | r | du Pont |
| 5633.61076 | 5.044 | 0.052 | 0.855 | 43.501 | 42.747 | 1.57536 | 21.391 | r | du Pont |
| 5633.61505 | 5.185 | 0.052 | 0.855 | 43.501 | 42.748 | 1.57680 | 21.532 | r | du Pont |
| 5633.61932 | 5.176 | 0.052 | 0.855 | 43.501 | 42.748 | 1.57680 | 21.523 | r | du Pont |
| 5893.75072 | 5.366 | 0.216 | 1.212 | 43.498 | 43.125 | 4.71888 | 21.732 | R* | SMARTS |
| 5893.83871 | 5.463 | 0.148 | 1.211 | 43.498 | 43.124 | 4.70736 | 21.829 | R* | SMARTS |
| 5894.74812 | 4.934 | 0.166 | 1.203 | 43.498 | 43.109 | 4.58640 | 21.299 | R | SMARTS |
| 5894.79508 | 5.155 | 0.105 | 1.203 | 43.498 | 43.109 | 4.57920 | 21.520 | R | SMARTS |
| 5894.82387 | 5.349 | 0.132 | 1.202 | 43.498 | 43.108 | 4.57632 | 21.714 | R | SMARTS |
| 6004.51965 | 5.271 | 0.094 | 0.930 | 43.497 | 42.793 | 1.95696 | 21.620 | $r_B$ | du Pont |
| 6004.52392 | 5.239 | 0.094 | 0.930 | 43.497 | 42.793 | 1.95696 | 21.588 | $r_B$ | du Pont |
| 6004.52818 | 5.212 | 0.094 | 0.931 | 43.497 | 42.793 | 1.95696 | 21.561 | $r_B$ | du Pont |
| 6004.60891 | 5.090 | 0.094 | 0.932 | 43.497 | 42.794 | 1.96560 | 21.439 | $r_B$ | du Pont |
| 6004.61317 | 5.139 | 0.094 | 0.932 | 43.497 | 42.794 | 1.96560 | 21.488 | $r_B$ | du Pont |
| 6004.61744 | 4.996 | 0.094 | 0.932 | 43.497 | 42.794 | 1.96560 | 21.345 | $r_B$ | du Pont |
| 6005.50917 | 5.291 | 0.020 | 0.947 | 43.497 | 42.805 | 2.05632 | 21.641 | $r_B$ | du Pont |
| 6005.51344 | 5.335 | 0.020 | 0.947 | 43.497 | 42.805 | 2.05776 | 21.685 | $r_B$ | du Pont |
| 6005.51771 | 5.312 | 0.021 | 0.947 | 43.497 | 42.805 | 2.05776 | 21.662 | $r_B$ | du Pont |
| 6005.65166 | 5.294 | 0.017 | 0.949 | 43.497 | 42.807 | 2.07072 | 21.644 | $r_B$ | du Pont |
| 6005.65593 | 5.345 | 0.020 | 0.949 | 43.497 | 42.807 | 2.07216 | 21.695 | $r_B$ | du Pont |
| 6005.66020 | 5.315 | 0.019 | 0.949 | 43.497 | 42.807 | 2.07216 | 21.665 | $r_B$ | du Pont |
| 6006.51153 | 5.679 | 0.129 | 0.963 | 43.497 | 42.818 | 2.16000 | 22.029 | $r_B$* | du Pont |
| 6006.51581 | 5.627 | 0.129 | 0.963 | 43.497 | 42.818 | 2.16000 | 21.977 | $r_B$* | du Pont |
| 6006.52009 | 5.568 | 0.129 | 0.963 | 43.497 | 42.818 | 2.16144 | 21.918 | $r_B$* | du Pont |
| 6006.52444 | 5.518 | 0.129 | 0.963 | 43.497 | 42.818 | 2.16144 | 21.868 | $r_B$* | du Pont |
| 6006.52872 | 5.562 | 0.129 | 0.963 | 43.497 | 42.818 | 2.16144 | 21.912 | $r_B$* | du Pont |
| 6006.53300 | 5.492 | 0.129 | 0.963 | 43.497 | 42.818 | 2.16144 | 21.842 | $r_B$* | du Pont |
| 6006.59374 | 5.363 | 0.129 | 0.964 | 43.497 | 42.819 | 2.16864 | 21.713 | $r_B$* | du Pont |
| 6006.59801 | 5.298 | 0.129 | 0.964 | 43.497 | 42.819 | 2.16864 | 21.648 | $r_B$* | du Pont |
| 6006.60229 | 5.311 | 0.129 | 0.964 | 43.497 | 42.819 | 2.16864 | 21.661 | $r_B$* | du Pont |
| 6007.51029 | 5.307 | 0.031 | 0.978 | 43.497 | 42.830 | 2.26512 | 21.658 | $r_B$ | du Pont |
| 6007.51456 | 5.284 | 0.031 | 0.978 | 43.497 | 42.830 | 2.26512 | 21.635 | $r_B$ | du Pont |
| 6007.51883 | 5.292 | 0.031 | 0.979 | 43.497 | 42.830 | 2.26512 | 21.643 | $r_B$ | du Pont |
| 6007.52318 | 5.280 | 0.031 | 0.979 | 43.497 | 42.830 | 2.26656 | 21.631 | $r_B$ | du Pont |
| 6007.52745 | 5.299 | 0.031 | 0.979 | 43.497 | 42.831 | 2.26656 | 21.650 | $r_B$ | du Pont |
| 6007.53172 | 5.346 | 0.031 | 0.979 | 43.497 | 42.831 | 2.26656 | 21.697 | $r_B$ | du Pont |
| 6007.59674 | 5.305 | 0.031 | 0.980 | 43.497 | 42.831 | 2.27376 | 21.656 | $r_B$ | du Pont |
| 6007.60101 | 5.222 | 0.031 | 0.980 | 43.497 | 42.831 | 2.27376 | 21.573 | $r_B$ | du Pont |
| 6007.60528 | 5.278 | 0.031 | 0.980 | 43.497 | 42.831 | 2.27520 | 21.629 | $r_B$ | du Pont |
| 6008.50809 | 5.190 | 0.047 | 0.994 | 43.497 | 42.843 | 2.37024 | 21.542 | $r_B$ | du Pont |
| 6008.51237 | 5.179 | 0.047 | 0.994 | 43.497 | 42.843 | 2.37168 | 21.531 | $r_B$ | du Pont |
| 6008.51666 | 5.239 | 0.047 | 0.994 | 43.497 | 42.843 | 2.37168 | 21.591 | $r_B$ | du Pont |
| 6008.52104 | 5.178 | 0.047 | 0.994 | 43.497 | 42.843 | 2.37168 | 21.530 | $r_B$ | du Pont |
| 6008.52532 | 5.214 | 0.047 | 0.994 | 43.497 | 42.843 | 2.37312 | 21.566 | $r_B$ | du Pont |
| 6008.52962 | 5.266 | 0.047 | 0.994 | 43.497 | 42.843 | 2.37312 | 21.618 | $r_B$ | du Pont |
| 6008.58998 | 5.183 | 0.047 | 0.995 | 43.497 | 42.844 | 2.37888 | 21.535 | $r_B$ | du Pont |
| 6008.59426 | 5.285 | 0.047 | 0.995 | 43.497 | 42.844 | 2.38032 | 21.637 | $r_B$ | du Pont |
| 6008.59854 | 5.306 | 0.047 | 0.995 | 43.497 | 42.844 | 2.38032 | 21.658 | $r_B$ | du Pont |
| 6274.06823 | 6.229 | 0.021 | 1.058 | 43.493 | 42.911 | 2.93760 | 22.584 | g | Gemini |
| 6274.07214 | 6.247 | 0.022 | 1.058 | 43.493 | 42.911 | 2.93760 | 22.602 | g | Gemini |
| 6274.07551 | 4.918 | 0.044 | 1.057 | 43.493 | 42.911 | 2.93616 | 21.273 | i | Gemini |
| 6274.07825 | 4.917 | 0.036 | 1.057 | 43.493 | 42.911 | 2.93616 | 21.272 | i | Gemini |
| 6274.08102 | 4.920 | 0.018 | 1.057 | 43.493 | 42.911 | 2.93616 | 21.275 | i | Gemini |
| 6276.08448 | 6.328 | 0.023 | 1.029 | 43.493 | 42.883 | 2.70432 | 22.682 | g | Gemini |
| 6276.08840 | 6.343 | 0.028 | 1.029 | 43.493 | 42.883 | 2.70432 | 22.697 | g | Gemini |
| 6276.09177 | 4.988 | 0.016 | 1.029 | 43.493 | 42.883 | 2.70288 | 21.341 | i | Gemini |
| 6276.09453 | 4.982 | 0.016 | 1.029 | 43.493 | 42.883 | 2.70288 | 21.335 | i | Gemini |
| 6276.09730 | 4.988 | 0.016 | 1.029 | 43.493 | 42.883 | 2.70288 | 21.341 | i | Gemini |
| 6277.07493 | 6.237 | 0.023 | 1.015 | 43.493 | 42.870 | 2.59200 | 22.590 | g | Gemini |
| 6277.07884 | 6.232 | 0.021 | 1.015 | 43.493 | 42.870 | 2.59200 | 22.585 | g | Gemini |
| 6277.08221 | 4.934 | 0.015 | 1.015 | 43.493 | 42.870 | 2.59056 | 21.287 | i | Gemini |
| 6277.08497 | 4.926 | 0.015 | 1.015 | 43.493 | 42.870 | 2.59056 | 21.279 | i | Gemini |
| 6277.08774 | 4.925 | 0.016 | 1.015 | 43.493 | 42.869 | 2.59056 | 21.278 | i | Gemini |
| 6278.09360 | 6.185 | 0.024 | 1.000 | 43.493 | 42.856 | 2.47824 | 22.538 | g | Gemini |
| 6278.09751 | 6.161 | 0.022 | 1.000 | 43.493 | 42.856 | 2.47824 | 22.514 | g | Gemini |
| 6278.10088 | 4.883 | 0.016 | 1.000 | 43.493 | 42.856 | 2.47680 | 21.235 | i | Gemini |
| 6278.10363 | 4.919 | 0.017 | 1.000 | 43.493 | 42.856 | 2.47680 | 21.271 | i | Gemini |
| 6278.10641 | 4.907 | 0.017 | 1.000 | 43.493 | 42.856 | 2.47680 | 21.259 | i | Gemini |
| 6279.06835 | 6.268 | 0.023 | 0.985 | 43.493 | 42.843 | 2.37024 | 22.620 | g | Gemini |
| 6279.07226 | 6.233 | 0.023 | 0.985 | 43.493 | 42.843 | 2.37024 | 22.585 | g | Gemini |
| 6279.07563 | 4.922 | 0.029 | 0.985 | 43.493 | 42.843 | 2.37024 | 21.273 | i | Gemini |
| 6279.07838 | 4.897 | 0.030 | 0.985 | 43.493 | 42.843 | 2.37024 | 21.248 | i | Gemini |
| 6279.08116 | 4.932 | 0.026 | 0.985 | 43.493 | 42.843 | 2.36880 | 21.283 | i | Gemini |
| 6280.01725 | 6.250 | 0.024 | 0.971 | 43.493 | 42.831 | 2.26800 | 22.601 | g | Gemini |
| 6280.02116 | 6.277 | 0.025 | 0.971 | 43.493 | 42.831 | 2.26800 | 22.628 | g | Gemini |
| 6280.02453 | 4.893 | 0.028 | 0.971 | 43.493 | 42.831 | 2.26656 | 21.244 | i | Gemini |
| 6280.02728 | 4.911 | 0.031 | 0.971 | 43.493 | 42.831 | 2.26656 | 21.262 | i | Gemini |
| 6280.03007 | 4.895 | 0.028 | 0.971 | 43.493 | 42.831 | 2.26656 | 21.246 | i | Gemini |
| 6281.04185 | 6.319 | 0.024 | 0.955 | 43.493 | 42.818 | 2.15856 | 22.670 | g | Gemini |
| 6281.04576 | 6.331 | 0.027 | 0.955 | 43.493 | 42.817 | 2.15856 | 22.682 | g | Gemini |
| 6281.04913 | 4.961 | 0.027 | 0.955 | 43.493 | 42.817 | 2.15856 | 21.311 | i | Gemini |
| 6281.05189 | 4.973 | 0.032 | 0.955 | 43.493 | 42.817 | 2.15712 | 21.323 | i | Gemini |
| 6281.05466 | 4.977 | 0.027 | 0.954 | 43.493 | 42.817 | 2.15712 | 21.327 | i | Gemini |
| 6281.05883 | 6.289 | 0.021 | 0.954 | 43.493 | 42.817 | 2.15712 | 22.640 | g | Gemini |
| 6281.06220 | 4.961 | 0.029 | 0.954 | 43.493 | 42.817 | 2.15712 | 21.311 | i | Gemini |
| 6301.04482 | 6.221 | 0.035 | 0.583 | 43.493 | 42.610 | 0.43200 | 22.561 | g | Gemini |
| 6301.04873 | 6.253 | 0.034 | 0.583 | 43.493 | 42.610 | 0.43200 | 22.593 | g | Gemini |
| 6301.05208 | 4.872 | 0.038 | 0.583 | 43.493 | 42.610 | 0.43200 | 21.211 | i | Gemini |
| 6301.05484 | 4.916 | 0.036 | 0.583 | 43.493 | 42.610 | 0.43200 | 21.255 | i | Gemini |
| 6301.05764 | 4.907 | 0.039 | 0.583 | 43.493 | 42.610 | 0.43200 | 21.246 | i | Gemini |
| 6303.00117 | 6.143 | 0.030 | 0.542 | 43.493 | 42.595 | 0.30960 | 22.482 | g | Gemini |
| 6303.00510 | 6.121 | 0.041 | 0.542 | 43.493 | 42.595 | 0.30960 | 22.460 | g | Gemini |
| 6303.00846 | 4.769 | 0.039 | 0.541 | 43.493 | 42.595 | 0.30960 | 21.108 | i | Gemini |
| 6303.01122 | 4.799 | 0.041 | 0.541 | 43.493 | 42.595 | 0.30960 | 21.138 | i | Gemini |
| 6303.01400 | 4.808 | 0.042 | 0.541 | 43.493 | 42.595 | 0.30960 | 21.147 | i | Gemini |
| 6303.89633 | 6.121 | 0.031 | 0.522 | 43.493 | 42.589 | 0.25632 | 22.460 | g | Gemini |
| 6303.90026 | 6.134 | 0.034 | 0.522 | 43.493 | 42.589 | 0.25632 | 22.473 | g | Gemini |
| 6303.90361 | 4.789 | 0.037 | 0.522 | 43.493 | 42.589 | 0.25632 | 21.127 | i | Gemini |
| 6303.90637 | 4.804 | 0.035 | 0.522 | 43.493 | 42.589 | 0.25632 | 21.142 | i | Gemini |
| 6303.90917 | 4.808 | 0.036 | 0.522 | 43.493 | 42.589 | 0.25632 | 21.146 | i | Gemini |
| 6305.03581 | 6.209 | 0.024 | 0.498 | 43.493 | 42.581 | 0.19152 | 22.548 | g | Gemini |
| 6305.03972 | 6.301 | 0.032 | 0.498 | 43.493 | 42.581 | 0.19152 | 22.640 | g | Gemini |
| 6305.04308 | 4.856 | 0.013 | 0.498 | 43.493 | 42.581 | 0.19152 | 21.194 | i | Gemini |
| 6305.04582 | 4.913 | 0.012 | 0.498 | 43.493 | 42.581 | 0.19152 | 21.251 | i | Gemini |
| 6305.04859 | 4.819 | 0.014 | 0.498 | 43.493 | 42.581 | 0.19152 | 21.157 | i | Gemini |
| 6305.97451 | 6.206 | 0.032 | 0.478 | 43.493 | 42.575 | 0.14112 | 22.544 | g | Gemini |
| 6308.98012 | 6.126 | 0.023 | 0.411 | 43.493 | 42.557 | -0.00720 | 22.463 | g | Gemini |
| 6308.98402 | 6.127 | 0.025 | 0.411 | 43.493 | 42.557 | -0.00864 | 22.464 | g | Gemini |
| 6309.93513 | 6.122 | 0.045 | 0.390 | 43.493 | 42.552 | -0.05040 | 22.459 | g | Gemini |
| 6309.93903 | 6.161 | 0.037 | 0.390 | 43.493 | 42.552 | -0.05040 | 22.498 | g | Gemini |
| 6309.94241 | 4.783 | 0.019 | 0.390 | 43.493 | 42.552 | -0.05040 | 21.119 | i | Gemini |
| 6309.94514 | 4.794 | 0.019 | 0.390 | 43.493 | 42.552 | -0.05040 | 21.130 | i | Gemini |
| 6309.94788 | 4.764 | 0.016 | 0.390 | 43.493 | 42.552 | -0.05040 | 21.100 | i | Gemini |
| 6310.98179 | 6.059 | 0.027 | 0.367 | 43.493 | 42.547 | -0.09504 | 22.396 | g | Gemini |
| 6310.98569 | 6.128 | 0.044 | 0.366 | 43.493 | 42.547 | -0.09504 | 22.465 | g | Gemini |
| 6310.98906 | 4.758 | 0.021 | 0.366 | 43.493 | 42.547 | -0.09504 | 21.094 | i | Gemini |
| 6310.99182 | 4.789 | 0.019 | 0.366 | 43.493 | 42.547 | -0.09504 | 21.125 | i | Gemini |
| 6310.99455 | 4.788 | 0.018 | 0.366 | 43.493 | 42.547 | -0.09504 | 21.124 | i | Gemini |
| 6313.03858 | 6.266 | 0.088 | 0.320 | 43.493 | 42.537 | -0.17424 | 22.602 | g | Gemini |
| 6313.04586 | 4.750 | 0.110 | 0.320 | 43.493 | 42.537 | -0.17424 | 21.086 | i | Gemini |
| 6313.05137 | 4.829 | 0.109 | 0.320 | 43.493 | 42.537 | -0.17424 | 21.165 | i | Gemini |

**Table 4. Best-Fit Single-Peaked Light Curve Parameters**

| P (d) | $\sigma_P$(d) | $\beta$ (mag deg$^{-1}$) | $\sigma_\beta$ (mag deg$^{-1}$) | A(mag) | $\sigma_A$(mag) | $\phi$ | $\sigma_\phi$ | $\chi^2$ | N | $\nu$ | L |
|---|---|---|---|---|---|---|---|---|---|---|---|
| 6.2562 | 0.002 | 0.16 | 0.015 | 0.055 | 0.005 | 0.310 | 0.040 | 45.98 | 40 | 32 | 0.052 |
| 6.2550 | … | 0.16 | 0.015 | 0.055 | 0.005 | 0.287 | 0.015 | 46.35 | 40 | 33 | 0.061 |

**Table 5. Best-Fit Double-Peaked Light Curve Parameters**

| P (d) | $\beta$ (mag deg$^{-1}$) | $\sigma_\beta$ (mag deg$^{-1}$) | $A_o$ (mag) | $\sigma_{Ao}$ (mag) | $\phi_o$ | $\sigma_{\phi o}$ | $A_\Delta$ (mag) | $\sigma_{A\Delta}$ (mag) | $\phi_\Delta$ | $\sigma_{\phi\Delta}$ | $\chi^2$ | N | $\nu$ | L |
|---|---|---|---|---|---|---|---|---|---|---|---|---|---|---|
| 12.510 | 0.147 | 0.018 | 0.060 | 0.006 | 0.265 | 0.015 | 0.022 | 0.005 | 0.87 | 0.06 | 33.30 | 40 | 31 | 0.36 |

**Table 6. Chi-Square Fits to Solar Phase Curves**

| Data Set | N | H(mag) | $\sigma_H$(mag) | $\chi^2$ | L |
|---|---|---|---|---|---|
| SMARTS R | 8 | 4.918 | 0.031 | 8.11 | 0.32 |
| du Pont r | 7 | 5.114 | 0.008 | 1.5 | 0.95 |
| Gemini g | 13 | 6.1 | 0.006 | 16.67 | 0.16 |
| Gemini i | 12 | 4.771 | 0.006 | 7.94 | 0.71 |

**Figure 1**. The minimum $\chi^2$ for a single-peaked sinusoidal fit to the rotation light curve of Sila-Nunam versus rotation period: (a) for the full range of explored periods; (b) for a narrow range of periods near the best-fit value. In (b), the dashed vertical line marks the period $P = P_{orb}/2 = 6.2550$ d, where $P_{orb}$ is the binary orbital period (Benecchi et al., 2013). The best fit from the light curve fit occurs at $P = 6.2562$ d, with dashed horizontal lines marking the 68.3% and 95.4% confidence limits. These respectively correspond to uncertainty ranges -0.001 to +0.002 d and -0.004 to + 0.005 d with respect to the minimum. The best-fit period is consistent with $P_{orb}/2$ at the ~1-sigma confidence level.

**Figure 2.** Reduced magnitude versus rotational phase for Sila-Nunam as measured (from top to bottom) in Gunn r with the du Pont telescope (small, filled red squares); in Johnson R with SMARTS (large, unfilled blue squares); in Gunn i with Gemini telescope (black triangles); and in Gunn g with the Gemini telescope (large, unfilled green circles). All observations have been corrected for solar phase angle dependence. Each band has been arbitrarily shifted in the vertical direction for clarity. Rotational phase is calculated assuming a rotational period, $P = 12.50995$ d, matching the binary orbital period. Black curves show the best-fit, double-peaked sinusoid with this same period, simultaneously fit to all the observations (see text for details). All observations are nightly averages except for those du Pont and SMARTS observation obtained during eclipsing events (these appear at rotational phase < 0.03).

**Figure 3.** Reduced magnitude (a) and residual reduced magnitude (b) versus rotational phase for Sila-Nunam. These are the same, solar-phase corrected data points plotted in Figure 2, but with the best-fit values for H in each band pass (see Table 5) subtracted from the respective magnitudes. This normalizes all the observations to the same average brightness. Under our assumption that there is no color dependence to the rotational modulation, all the observations should combine to yield a consistent light curve. The black curve shows the best-fit double-peaked light curve, $F_d(t)$, which has been subtracted to yield the residual magnitudes plotted in (b). See Figure 2 for symbol explanations.

**Figure 4**. Reduced magnitudes versus rotational phase in Johnson R (unfilled blue triangles) and Gunn r(filled red squares) after subtracting the best-fit solar phase dependence and double-peaked sinusoid with $P = P_{orb}$ (see Fig 2). Only the observations coinciding with predicted mutual events are shown, with nearly all occurring near zero rotational phase (i.e. near the light curve minimum). For the Johnson R band, all the eclipse observations but one coincide with an event near JD= 2455593 (a single observation not shown occurs at rotational phase -0.508, coinciding with an event on JD 2455599). For the Gunn r band, all the observations coincide with a single event on JD 2456006. The expected eclipse modulation, assuming Lambertian-scattering spheres for Sila and Nunam, is shown by the solid blue and dotted red curves (for JD 2455593 and 2456006, respectively).

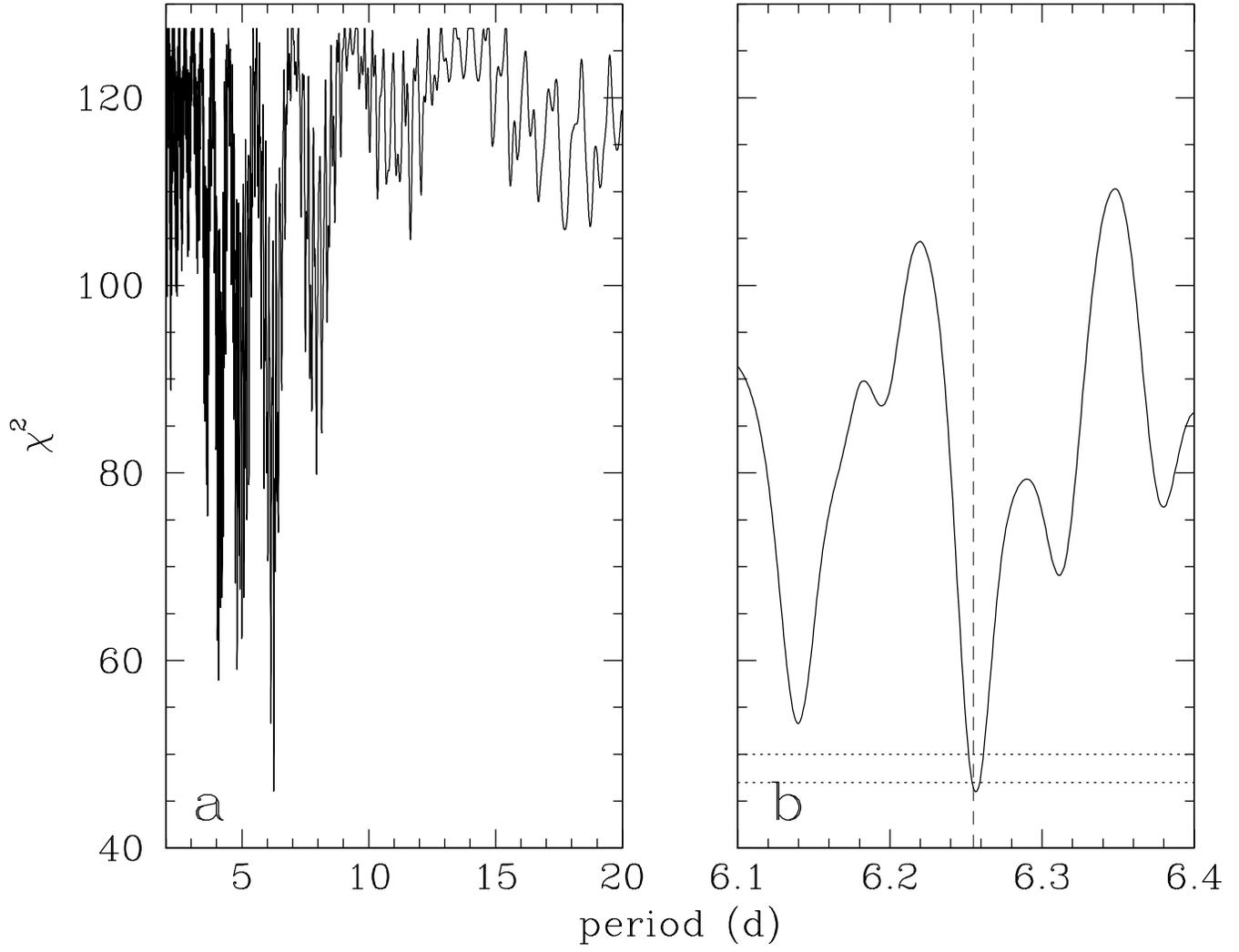

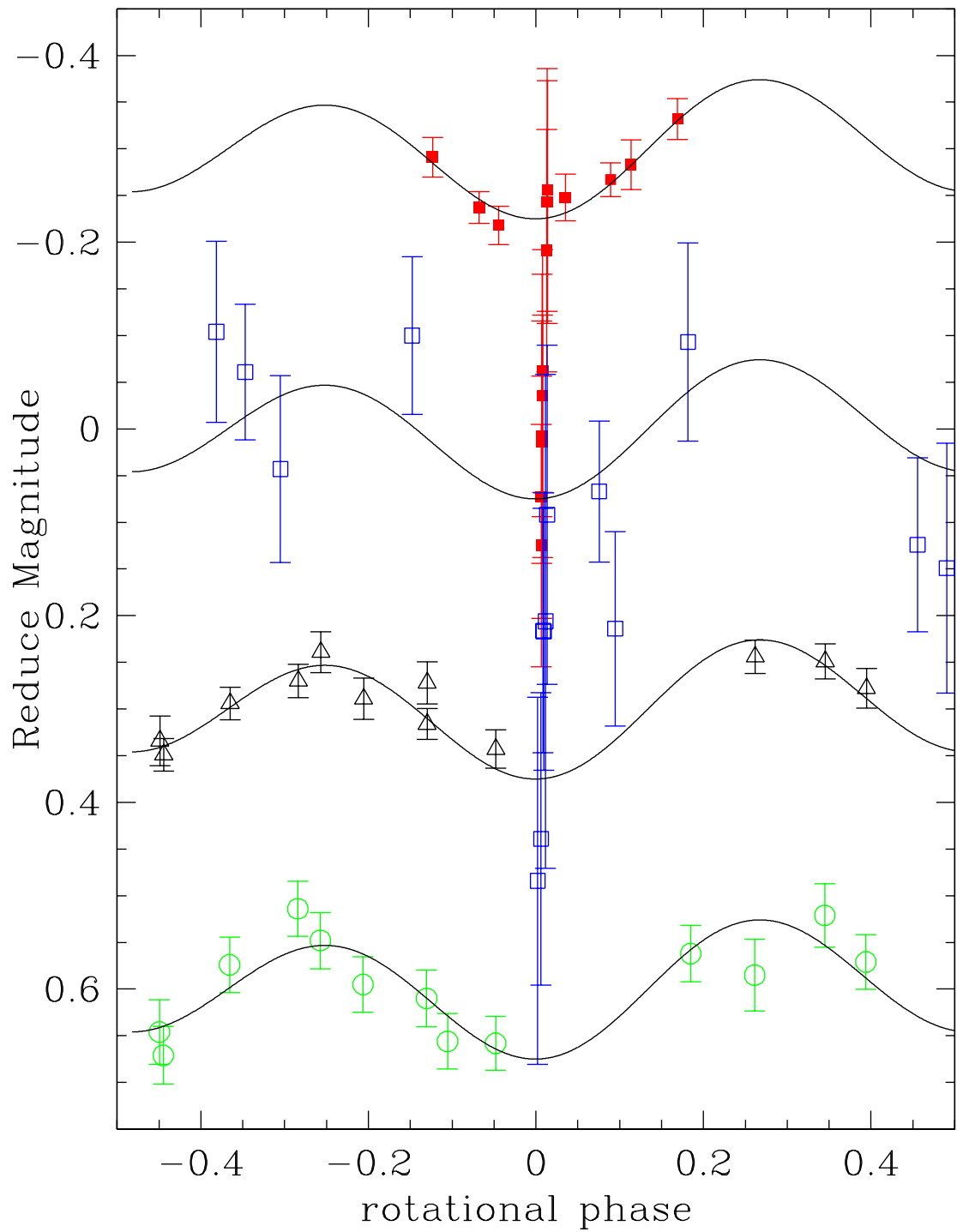

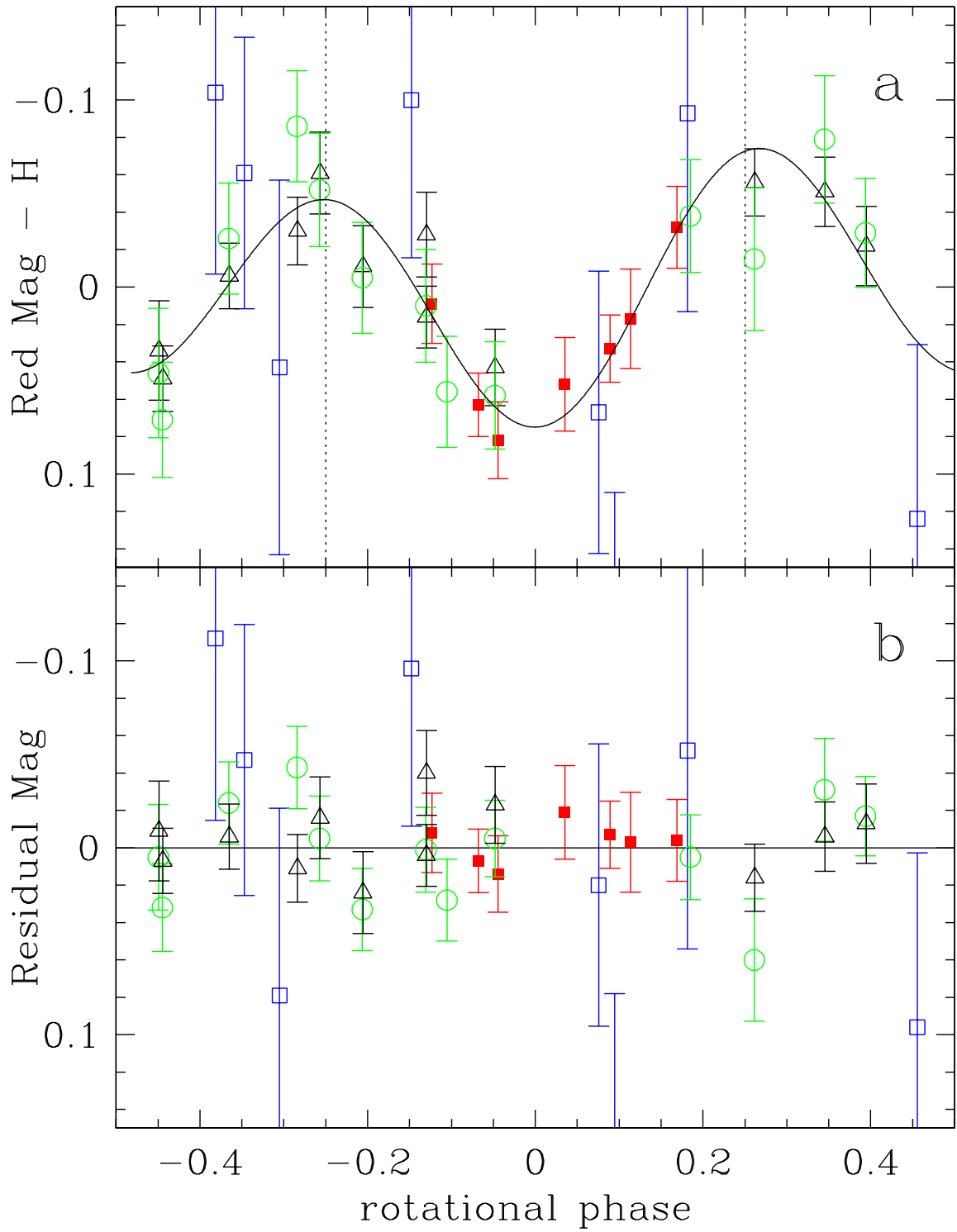

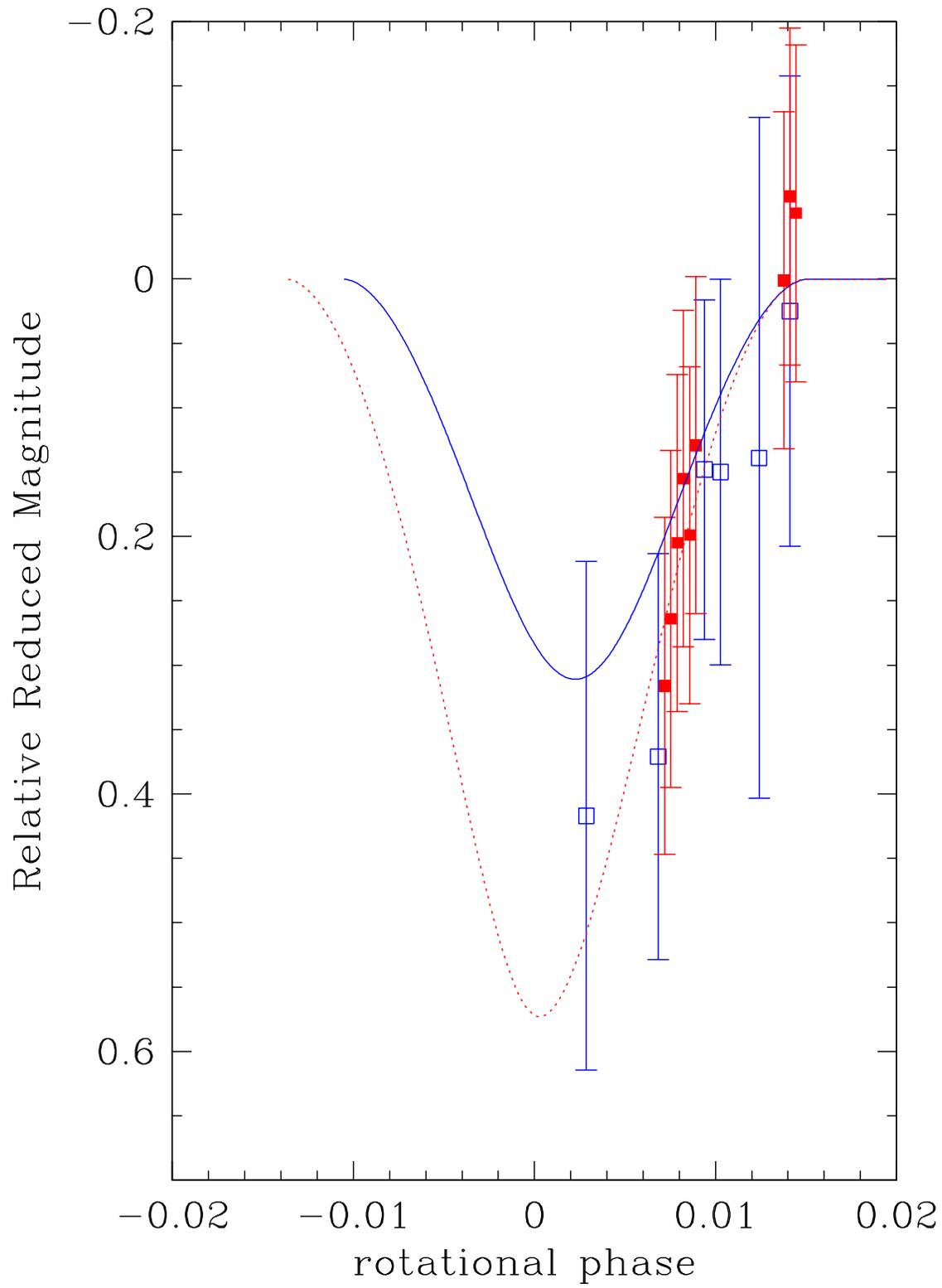